\documentclass[11pt,dvips]{article}

\usepackage{epsfig,times} 
\usepackage{picinpar}
\usepackage{wrapfig}
\usepackage{floatflt}
\makeatletter
\renewcommand{\section}{\@startsection{section}{1}{0in}
	{0.4\baselineskip}{0.1\baselineskip}{\Large\bf}}
\renewcommand{\subsection}{\@startsection{subsection}{2}{0in}
	{0.25\baselineskip}{-\baselineskip}{\large\bf}}
\renewcommand{\subsubsection}{\@startsection{subsubsection}{3}{0in}
	{0.1\baselineskip}{-\baselineskip}{\normalsize\bf}}
\makeatother
\def \deg       {$^o$ }
\def \dg       {$^o$}
\def \gray     {$\gamma$-ray }
\def \grays    {$\gamma$-rays }

\def \sig      {$\sigma$ }

\def\repeat{\hbox to40pt{\hrulefill}}

\def\bbllx{ 2.0cm}
\def\bblly{ 8.5cm}
\def\bburx{19.5cm}
\def\bbury{25.9cm}

\begin{document}

%
\makeatletter\newcommand{\ps@icrc}{
\renewcommand{\@oddhead}{\slshape{OG 2.1.22}\hfil}}
\makeatother\thispagestyle{icrc}
%
%

\begin{center}
{\LARGE \bf COMPTEL MeV Observations of the TeV Sources Markarian~421 and Markarian~501}
\end{center}

\begin{center}
%
%
{\bf W. Collmar$^{1}$, K. Bennett$^{4}$, H. Bloemen$^{2}$, W. Hermsen$^{2}$,
J. Ryan$^{3}$,\\
V. Sch\"onfelder$^{1}$, H. Steinle$^{1}$, O.R. Williams$^{4}$}\\
{\it $^{1}$Max-Planck-Institut f\"ur extraterrestrische Physik, P.O. Box 1603, 85740 Garching, Germany\\
     $^{2}$SRON-Utrecht, Sorbonnelaan 2, 3584 CA Utrecht, The Netherlands\\
     $^{3}$University of New Hampshire,  ISEOS, Durham NH 03824, USA\\
     $^{4}$Astrophysics Division, ESA/ESTEC, NL-2200 AG Noordwijk, The Netherlands}
\end{center}

\begin{center}
{\large \bf Abstract\\}
\end{center}
\vspace{-0.5ex}
%
%
The COMPTEL experiment aboard the COMPTON Gamma-Ray Observatory (CGRO) 
has observed the prominent TeV-blazars Mkn~421 and Mkn~501 many times 
between the start of its mission in April '91 and December '98.
This paper reports first COMPTEL results from mainly time-averaged 
(CGRO Cycles) data.
No evidence for both sources is found up to the end Cycle VI.  
However, the sum of all 10-30~MeV Cycle VII data 
shows a weak (3.2\sig detection) MeV-source being positionally consistent
with Mkn~421. During Cycle VII Mkn~421 was rather active at TeV-energies.  
Due to the lack of other known \gray sources in this sky region, we consider
Mkn~421 as the most likely counterpart for this \gray emission.
However, its connection cannot be proven by COMPTEL. 
%

\vspace{1ex}

%
%
\section{Introduction:}
\label{intro.sec}
The discovery by the CGRO experiments
that blazars sometimes radiate a large or even the major fraction of their luminosity at \gray energies marked a milestone in our knowledge on these
powerful sources.
Now, after 8 years of operation, roughly 80 blazars have been detected
by EGRET at \gray energies above $\sim$100~MeV (e.g. Hartman et al. 1999).   COMPTEL, covering the soft \gray range (0.75-30~MeV) has detected 9 of these EGRET blazars (Collmar et al. 1999).

During recent years some blazars have been discovered to emit
even at TeV-energies by
the Whipple Observatory: the blazars Mkn~421 (Punch et al. 1992), Mkn~501 (Quinn et al. 1996), and 1~ES~2344+514 (Catanese et al. 1998). The most prominent ones are Mkn~421 and Mkn~501, which have been
detected many times by Whipple and confirmed by other TeV-experiments 
like the HEGRA Cherenkov telescopes (e.g. Aharonian et al. 1999).
Despite their (occasional) prominence at TeV-energies
-- during flaring periods Mkn~501 was the brightest TeV-source in the sky --
they are weak EGRET sources. Mkn~421 shows a weak flux in the 
EGRET band and Mkn~501, at the time of its TeV-discovery, was not detected
at all by EGRET. 
A common feature of all TeV-blazars is their extreme variability
on time-scales of years, days, hours, or even minutes.

COMPTEL, along the course of its now 8-year mission, has had the prominent 
TeV-blazars Mkn~421 and Mkn~501 several times within its field-of-view. The 
complete set of data (April '91 to Nov. '98) on these sources has been 
analysed. In this paper we will concentrate on time-averaged results, i.e. 
results of combined data over periods of typically one year.

\section{Observations and Data Analysis:}
\label{obs.sec}

CGRO observations are organized in so called Cycles 
(or Phases) which last for a time period of roughly one year, consisting
of many (30 - 40) individual pointings (called Viewing Periods: VPs), which 
typically last for two weeks. Table~1 provides the COMPTEL exposures 
on both sources for the individual CGRO cycles. It clearly shows that Mkn~421 was favorably located for COMPTEL in Cycle~VII. This is due to its proximity ($\sim$25\dg) to SN~1998bu which was a major COMPTEL target in 1998.   

\begin{table}[th]
\caption{COMPTEL exposures on Mkn~421 and Mkn~501 in individual CGRO cycles.
 The table provides the time periods of the individual cycles, and for both sources  the number of days within 30\deg and 20\deg of the COMPTEL pointing direction
and the effective exposures (net observation time with 100\% COMPTEL
efficiency).  }\label{Obs}
\vspace{12pt}
\begin{tabular}{|cccccc|}\hline
 CGRO        & Time Intervals & \multicolumn{2}{c}{Mkn~421}  &  \multicolumn{2}{c|}{Mkn~501}  \\
Phase/Cycle  & yy/mm/dd - yy/mm/dd  & [Days $<$30\dg/20\deg ] & Eff. Exp.  
& [Days $<$30\dg/20\deg ] & Eff. Exp.  \\
\hline
I    & 91/05/16 - 92/11/17 & 35/14  &  8.19  &  9/9  &  5.76 \\
II   & 92/11/17 - 93/08/17 & 48/21  &  6.61  & 14/14 &  3.49 \\
III  & 93/08/17 - 94/10/04 & 21/21  & 11.01  &  0/0  &  2.66 \\
IV   & 94/10/04 - 95/10/03 & 14/14  &  3.96  &  8/8  &  1.89 \\
V    & 95/10/03 - 96/10/15 & 14/14  &  2.37  & 27/27 & 10.06 \\
VI   & 96/10/15 - 97/11/11 &  0/0   &  0.00  & 28/6  &  7.18 \\
VII  & 97/11/11 - 98/12/01 & 108/25 & 22.43  & 23/9  & 10.39 \\
\hline
\end{tabular}\end{table}

We have applied the standard COMPTEL maximum-likelihood analysis method
(e.g. de Boer et al. 1992) to derive detection significances, fluxes, and 
flux errors of \gray sources in the four standard COMPTEL energy bands (0.75-1~MeV, 1-3~MeV, 3-10~MeV, 10-30~MeV), and a background modelling
technique which eliminates any source signature but preserves the general background structure (Bloemen et al. 1994). 
The source fluxes for both sources have been derived by a flux fitting 
procedure which iteratively determines the flux of one or more 
potential sources and simultaneously a background model which takes
into account the presence of possible sources.
As there are no known nearby \gray sources or source candidates,
for the case of Mkn~421 no other source was included in this procedure,  
while for the case of Mkn~501 the prominent EGRET quasar 4C+38.41,
which is only $\sim$4\deg away, was included as a second source. 
Because Mkn~421 (l/b: 179.8/65.0) and Mkn~501 (l/b: 63.6/38.9) are located
at high galactic latitudes, 
the present analyses have been carried out consistently in local coordinate
systems, i.e. centered on each source. 

\section{Results:}
\label{results.sec}

\subsection{Markarian~421:}
Up to the end of CGRO Cycle VI (November '97) no convincing evidence for
Mkn~421 could be found in any of the four standard COMPTEL energy bands in
the different CGRO cycles. However, in the 10-30~MeV map of CGRO Cycle VII 
a source-like feature appears which is positionally consistent with Mkn~421 (Fig.~1). The detection significance at the position of Mkn~421 is formally 3.2\sig assuming $\chi^2_1$-statistics for a known source, i.e. close to the detection threshold.
We checked the 3rd EGRET catalogue (Hartman et al. 1999) for \gray sources in this region. Apart from Mkn~421, the catalogue lists no other source which
could be responsible for this \gray emission. Therefore, we consider Mkn~421
as the most likely candidate. This is supported by the fact, that in 1998
Mkn~421 was unusually active at TeV-energies, sometimes even brigther than the Crab (Aharonian et al. 1999). However, a different origin for this \gray emission than Mkn~421 cannot 
be excluded. EGRET cannot help to resolve this issue, because it observed 
this sky position simultaneously only for 4 out of the 108 COMPTEL days (Table~1): either the pointings were too far off for its narrow-field-of-view mode or EGRET was switched off. 
%
%
\begin{figure} [tb]
\epsfig{figure=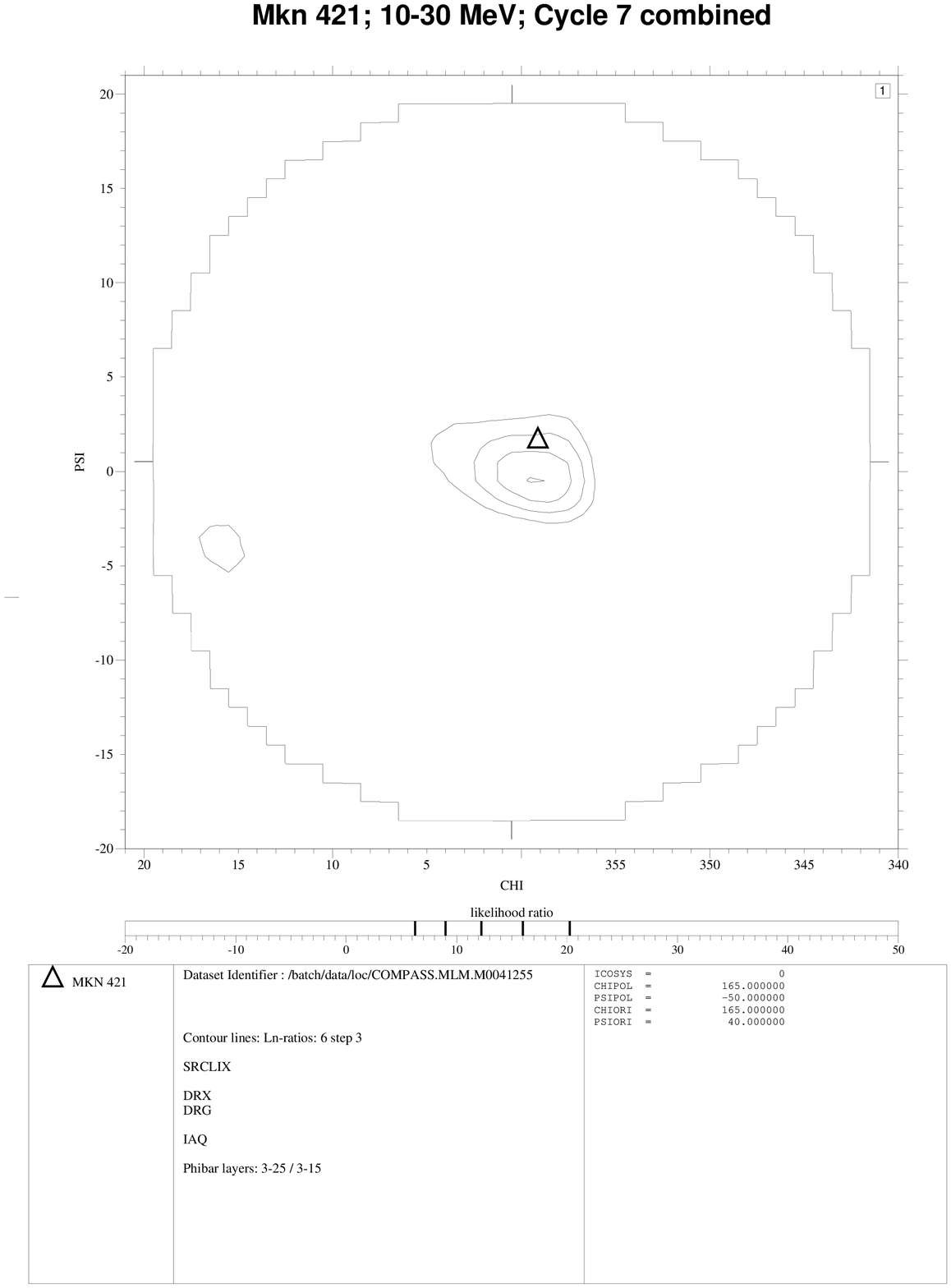,width=6.3cm,height=6.3cm,bbllx=\bbllx,bblly=\bblly,bburx=\bburx,bbury=\bbury,clip=}
\vspace*{-6.4cm}\hfill
\epsfig{figure=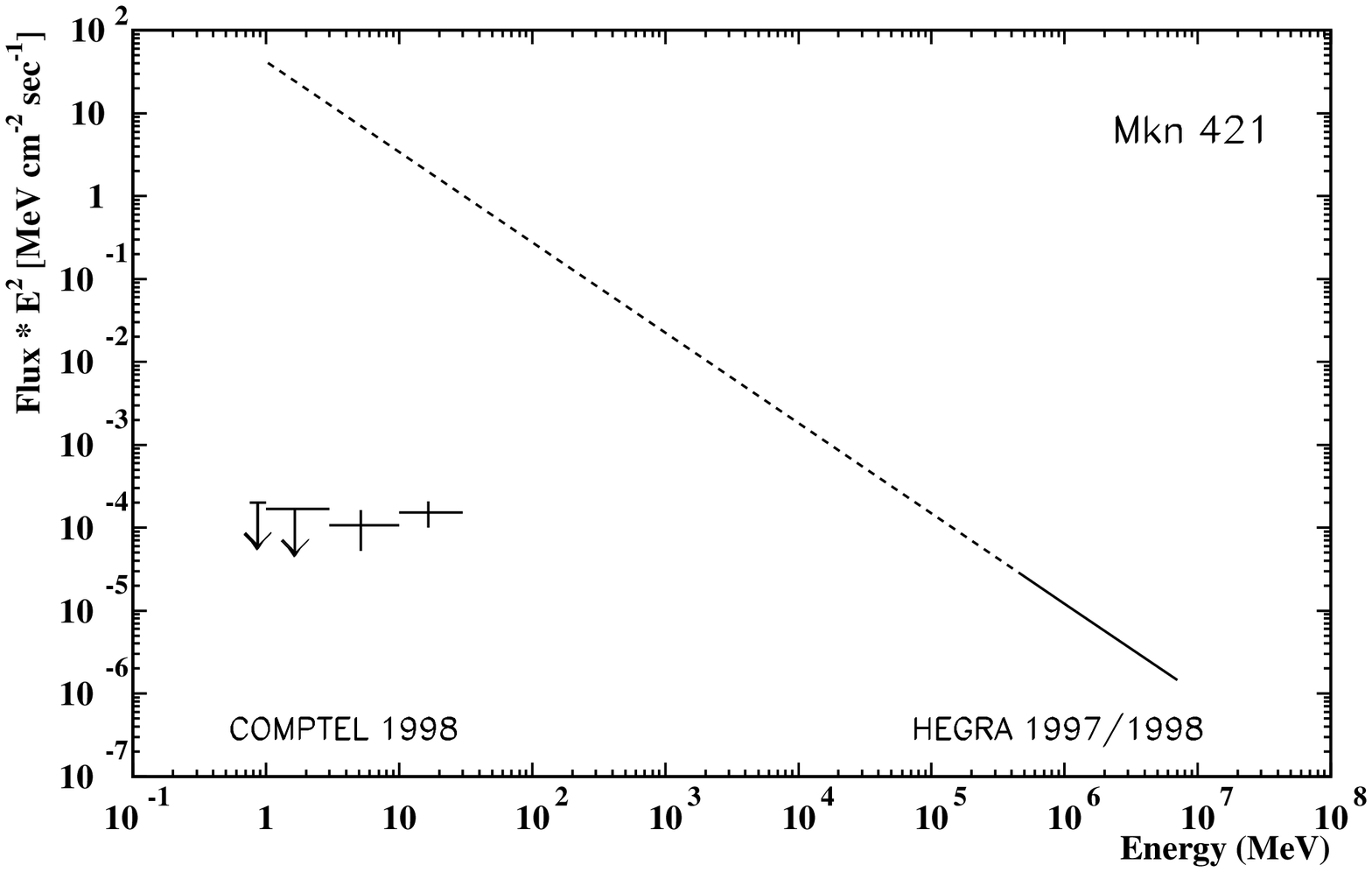,height=6.3cm,width=9.0cm,clip=}
\vspace{6.1cm}
\caption[]{Left: The 10-30~MeV maximimum likelihood skymap around the
position of
Mkn~421 ({\bf $\triangle$}) in local coordinates ($\chi$/$\psi$) for
the sum of all Cycle VII data. The contour lines start at a detection significance of
2.5$\sigma$ ($\chi^{2}_{1}$-statistics) with a step of 0.5$\sigma$.  
The \gray emission is positionally consistent with Mkn~421.
Right, comparison of the COMPTEL Cycle VII flux results to the 
average 1997-1998 HEGRA TeV-spectrum (Aharonian et al. 1999). The solid line shows the energy range (500~GeV to 7~TeV) of the HEGRA measurements, 
and the dashed line the spectral extrapolation of the reported 
power-law shape towards lower energies.}
\end{figure}
%
Applying the procedure described in Section~2, fluxes from the position 
of Mkn~421 have been derived. In addition to the 3\sig result in the 10-30~MeV
band of the Cycle VII data, there is a 2\sig flux point derived in the
3-10~MeV band for this time period.
Below 3~MeV the data are consistent with noise resulting in upper limits only. 
In Figure~1 these flux values are compared to the time-averaged 1997-1998 HEGRA TeV-spectrum (Aharonian et al. 1999).
The authors provide the spectral slope for the year 1998 individually ($\alpha$: -3.00$\pm$0.05) but not the flux normalisation. They note that the spectral slopes in both years are consistent within statistics.
The comparison shows that Mkn~421 (if responsible for the emission) 
radiated in 1998 on average more power at MeV-energies than at TeV-energies.  
In addition, the MeV spectral points are several orders of magnitudes below
the power-law extrapolation of the TeV-spectrum, proving that the TeV-spectrum
has to bend over above 30~MeV. The last conclusion is also valid of course,
if the observed \grays are not connected to Mkn~421. We like to note, that a 
large fraction of the COMPTEL Cycle VII exposure on Mkn~421 was collected 
during times (July to September 1998), when the source was inaccesible for TeV-observations.
%
%
\begin{figure} [tb]
\epsfig{figure=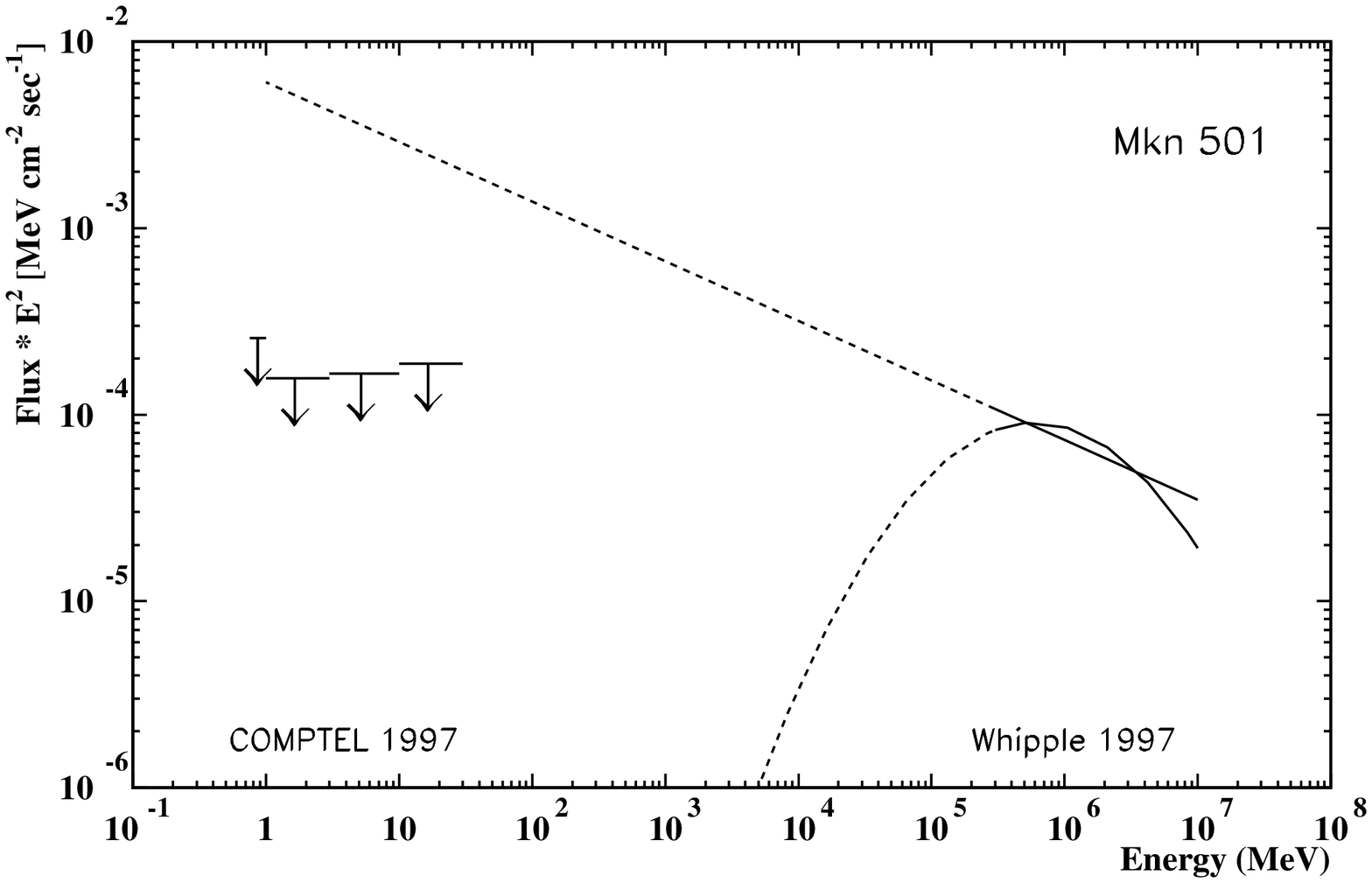,height=5.9cm,width=8.0cm,clip=}
\vspace*{-6.0cm}\hfill
\epsfig{figure=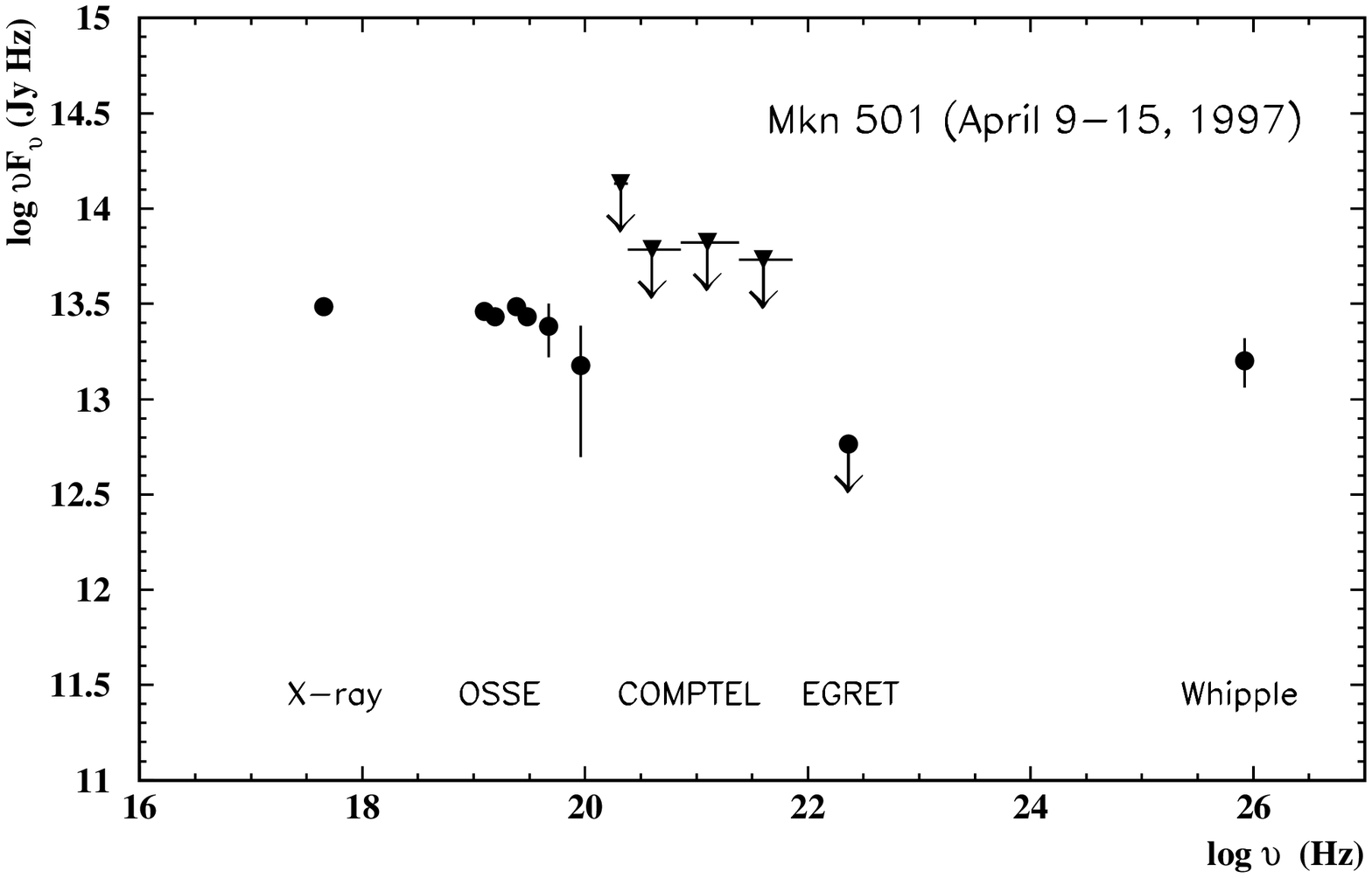,height=5.9cm,width=8.0cm,clip=}
\vspace{5.4cm}
\caption[]{Left: 
Comparison of the average COMPTEL Cycle VI upper limits of Mkn~501 with the 
quasi-simultaneous average 1997 TeV-spectrum as published  by the
Whipple collaboration (Samuelson et al. 1998). The solid lines show the energy range of the Whipple measurements (300~GeV - 10~TeV) and the dashed lines the spectral extrapolations for the power-law and curved spectral shape towards lower energies. Right, 
the COMPTEL upper limits are included in a simultaneously measured broad-band spectrum of Mkn~501 (Catanese et al. 1997).
The campaign was carried out (April 9-15, 1997) during an active TeV-period of Mkn~501. }
\end{figure}
%
\subsection{Markarian~501:}
Up to the end of CGRO Cycle VII (December~1998) no convincing evidence for
Mkn~501 could be found in any of the four standard COMPTEL energy
bands in the different CGRO Cycles. At TeV-energies Mkn~501 showed its largest
activity so far during the observational period in 1997 (e.g. Quinn et al. 
1999). To be quasi-simultaneous with that, we combined all COMPTEL 
Cycle VI data on Mkn~501 and derived the upper limits for its MeV flux. 
Unfortunately, only 6 days (a CGRO ToO during April 1997) of 
COMPTEL observations are directly simultaneous to the observed TeV-flaring period.
The COMPTEL 1997 upper limits are compared to the time-averaged 1997 
TeV-shape as observed by the Whipple telescope in Figure~2.
The COMPTEL upper limits are $\sim$2 orders of magnitude below the extrapolation
for an assumed power-law shape at TeV-energies, requiring spectral bending, but
still allowing for a luminosity in the MeV-band roughly equal to that detected 
in the TeV-band. Evidence for spectral bending 
was recently also found in the TeV-data itself (e.g. Samuelson et al. 1998).
The COMPTEL points are consistent with the extrapolation of that shape towards 
lower energies.       
As mentioned above, in April 1997 a multiwavelength campaign 
on Mkn~501 was carried out, where CGRO participated for 6 days (April 9-15, 1997) 
in a target-of-opportunity observation. COMPTEL has not detected the blazar
within this short observation period. The COMPTEL upper limits are shown 
in Figure~2, together with simultaneous flux measurements (Catanese et al. 1997)
from neighboring high-energy bands. The COMPTEL upper limits are consistent
with these measurements, 
however do not provide any further constraints.

\section{Summary and Conclusions:}
\label{summary.sec}
We present first COMPTEL MeV-results of the prominent TeV-blazars Mkn~421
and Mkn~501. So far, the analysis has been carried out in the four standard energy bands for individual CGRO VPs and individual CGRO cycles.
Up to the end of CGRO Cycle VI (November 1997), no evidence for Mkn~421 
was found. However, the combined Cycle VII 10-30~MeV data show evidence 
(although near the detection threshold) for a \gray source which is positionally
consistent with Mkn~421. As there are no other known \gray sources in that 
sky region, we consider the TeV-blazar as the most likely counterpart. 
If this \gray emission really originates from Mkn~421, this would be an interesting result. Broad-band spectra for flaring TeV-blazars indicate 
a spectral minimum near the COMPTEL and EGRET bands (e.g. Fig.~2), 
i.e. they should be 
located in the "spectral valley" between the peaks of the assumed synchrotron and inverse Compton (IC) emission components. This result would mean that 
either the synchrotron emission moved up to the highest energies ever observed, or the IC emission at MeV-energies was as high as never observed before for any TeV-blazar which would be the most likely explanation in our mind,
or something else (e.g. a combination of both) has happened.    
Eventually, multiwavelength spectra might help to resolve this issue. 
No convincing evidence for Mkn~501 is found in any of the analysed data. 
Derived upper limits, which are quasi-simultaneous with the
large TeV-flaring period in 1997, require a spectral bending
for a TeV power law spectrum, 
and are consistent with the extrapolation of a reported curved TeV-spectrum. COMPTEL participation in a multiwavelength campaign resulted
in upper limits which 
are consistent with the simultaneous measurements in the neighboring energy bands, however do not provide any further constraints.           
%
\vspace{1ex}
\begin{center}
{\Large\bf References}
\end{center}
Aharonian et al. 1999, A\&A submitted (astro-ph/9905032)\\
Bloemen, H., Hermsen, W., Swanenburg, B.N., et al. 1994, ApJ Suppl. 92, 419\\
de Boer, H., Bennett, K., Bloemen, H., et al. 1992, In: {\it Data Analysis in Astronomy IV}, eds. V. Di Gesu et al. (New York: plenum Press), 241\\
Catanese, M., et al. 1997, ApJ 487, L143\\ 
Catanese, M., et al. 1998, ApJ 501, 616\\ 
Collmar, W. et al., 1999, Proc. 3rd INTEGRAL Conf., in press\\  
Hartman, R. et al., 1999, ApJ Suppl., accepted\\
Punch, M. et al. 1992, Nature 358, 477\\
Quinn, J. et al. 1996, ApJ 456, L83\\
Quinn, J. et al. 1999, ApJ, submitted\\
Samuelson, F.W. et al. 1998, ApJ 501, L17\\

\end{document}